\documentclass[runningheads]{llncs}
\usepackage[T1]{fontenc}
\usepackage{graphicx}
\usepackage{hyperref}
\usepackage{braket}
\usepackage{mathtools}
\usepackage{tikz}
\usepackage{qcircuit}
\usepackage{multicol}

\usepackage{cite}
\usepackage[margin=3.5cm]{geometry}
\usepackage{amssymb,mathtools} 
\usepackage{amsmath} 
\usepackage{graphicx,placeins} 
\usepackage{ifthen}
\usepackage{pgfmath}
\usepackage[T1]{fontenc}
\usepackage{booktabs}         % prettier horizontal rules
\usepackage{hyperref}
\usepackage{amsmath}
\renewcommand{\vec}{\overrightarrow}
\usepackage{tikz}
\usetikzlibrary{calc,arrows}  
\usepackage{enumerate,comment}

\usepackage{float}
\usepackage{lipsum,xcolor,comment}

\usepackage{dcolumn}% Align table columns on decimal point
\usepackage{bm}% bold math
\usepackage{subcaption}
\begin{document}

\title{No Scratch Quantum Computing by Reducing Qubit Overhead for Efficient Arithmetics}

\author{Omid Faizy \inst{1,2,3,4} \and Norbert Wehn \inst{3} \and Paul Lukowicz \inst{4, 5} \and Maximilian Kiefer-Emmanouilidis \inst{4,5} }

\institute{Laboratoire de Chimie de la Matière Condensée de Paris, UMR CNRS 7574, Sorbonne Université, 4, place Jussieu, 75252 Paris, France. \\ \and
Laboratoire de Mathématiques Pures et Appliquées Joseph Liouville, Université du Littoral Côte d'Opale, 50 rue Ferdinand Buisson, CS 80699, 62228 Calais, France.\\ \and
 Microelectronic Systems Design (EMS), RPTU Kaiserslautern-Landau, Germany\\\and
 Department of Computer Science and Research Initiative QC-AI, RPTU Kaiserslautern-Landau, Kaiserslautern, Germany.\\ \and
 German Research Center for Artificial Intelligence (DFKI),
 Kaiserslautern, Germany.}

 \date{\today}
\maketitle

\begin{abstract}
Quantum arithmetic computation requires a substantial number of scratch qubits to stay reversible. These operations necessitate qubit and gate resources equivalent to those needed for the larger of the input or output registers due to state encoding. Quantum Hamiltonian Computing (QHC) introduces a novel approach by encoding input for logic operations within a single rotating quantum gate. This innovation reduces the required qubit register \( N \) to the size of the output states \( O \), where \( N = \log_2 O \). Leveraging QHC principles, we present reversible half-adder and full-adder circuits that compress the standard Toffoli + CNOT layout [Vedral et al., PRA, 54, 11, (1996)] %\cite{Vedral1996} 
from three-qubit and four-qubit formats for the Quantum half-adder circuit and five sequential Fredkin gates using five qubits [Moutinho et al., PRX Energy 2, 033002 (2023)] %\cite{Moutinho2023} 
for full-adder circuit; into a two-qubit, 4\(\times\)4 Hilbert space. This scheme, presented here, is optimized for classical logic evaluated on quantum hardware, which due to unitary evolution can bypass classical CMOS energy limitations to a certain degree. Although we avoid superposition of input and output states in this manuscript, this remains feasible in principle. We see the best application for QHC in finding the minimal qubit and gate resources needed to evaluate any truth table, advancing FPGA capabilities using integrated quantum circuits or photonics.
\end{abstract}

\keywords{Quantum Arithmetics  \and Quantum Hamiltonian Computing \and Adder.}

\section{Introduction}

The question of minimal and most efficient arithmetic operators is  approaching the atomic limits. Below a few hundred atoms per transistor, leakage and energy dissipation at least $kT\ln2$ per bit become prohibitive \cite{Landauer1961,bennett1973logical, Chiribella2021}.  Reversible quantum logic can bypass these bounds to a certain degree \cite{ Chiribella2021,bennett1973logical,Vedral1996}. Following this principle, early quantum adders \cite{Vedral1996,cuccaro2004new} implemented binary addition using Toffoli and CNOT gates, but at the expense of extra ancilla qubits for reversibility and state embedding. In this manuscript, we consider any qubit not needed to encode the output as an ancillary qubit. This distinguishes our approach from other ancilla-free methods which do not need any extra qubits when transpiled into universal gates, such as the three-qubit half adder, the four-qubit full adder. Here, the qubits necessary for input embedding are not treated as ancillas. To emphasize this distinction,
we refer to scratch qubits or scratch registers instead, which identifies qubits needed for the embedding, as a form of ancilla.

For instance, standard reversible half-adder and full-adder circuits \cite{Moutinho2023} require three and four qubits, respectively, in order to produce sum and carry outputs without erasing the inputs. This holds true even if the inputs are classical so that input and output are product states, and thus separable states. This "scratch" qubit overhead not only increases the Hilbert space (dimension 8 for three and dimension 16 for four qubits) but also adds numerous gates to uncompute intermediate results, complicating practical implementation. More recent designs reduce either the T-count or the ancilla count—sometimes to one or even zero \cite{Gidney2018, Remaud2025, remaud2024optimizing}, where additional ancillas refer to qubits necessary beyond the aforementioned input embedding for half-adders and four qubits for full-adders. 

Quantum Hamiltonian Computing (QHC) encodes Boolean inputs in Hamiltonian matrix elements rather than basis populations, minimizing the state count for logic gates \cite{Dridi_2015,Ample2017, Dridi2018,Namarvar2019}. Previous QHC studies implemented on single-molecule platforms realized 4-state half-adders and 8-state full-adders that required no qubits and relied on intrinsically irreversible dynamics; later work introduced multi-energy read-out schemes to compress the effective Hamiltonian even further \cite{Dridi2018,Namarvar2019}. 

Theoretical studies show that classical data can be encoded as rotation angles in a QHC circuit by using a nano-graphene molecule doubly functionalized with nitro groups and bridged by graphene electrodes, whose nitro-group rotations serve as binary inputs that collectively implement a half-adder Boolean operation
 \cite{SRIVASTAVA2017301}.
 
Adapting these ideas to quantum computing is to find a unitary $U(\vec{\theta})= \exp(-\mathrm{i}H(\vec{\theta})t)$ that depending on the binary inputs formulates the wanted logical output of any given truth table for the given input $\vec{\theta}$. From this point of view we argue that we can formulate a quantum representation (in the form of a quantum gate) which yields a given output (following the corresponding truth table) with only $N=\log_2(O)$ qubits where $O$ is the number of possible outputs.

Using the QHC paradigm, we present constructed reversible half- and full-adder circuits using only two qubits. This represents a significant reduction in the required resources compared to conventional designs. Both our half-adder and full-adder operate within a 4×4 Hilbert-space (two qubits), rather than the 8×8 (three qubit), or 16x16 (four qubits) or larger spaces required by previous methods. Currently, these circuits are restricted to classical Boolean inputs and cannot handle quantum superposition states at the input.

\section{Circuit design}
\begin{figure}[H]
 \centering
  \includegraphics[width=1\textwidth]{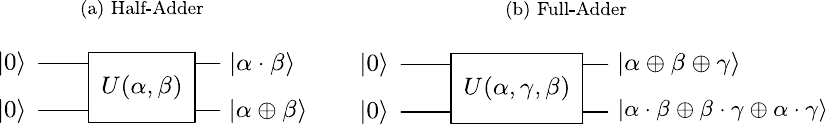}
     \caption{ (a) Two-Qubit QHC-Half-Adder circuit with classical binary inputs $\alpha$ and $\beta$ implemented directly inside circuit. (b) Two-Qubit QHC-Full-Adder circuit with classical binary inputs $\alpha$, $\gamma$ and $\beta$ implemented directly inside circuit.}
      \label{fig:2qubit_halfadder}
 \end{figure}

Starting with the half-adder as depicted in Fig.~\ref{fig:2qubit_halfadder} (a), the device employs two qubits prepared in $\ket{\psi(t=0)}=\ket{00}$.  A 4\,\(\times\)\,4 unitary ${U}(\alpha,\beta)$ confines dynamics to a protected sub‑manifold and 
encodes the angles $(\alpha,\beta)\in\{0,1\}^2$ as logical inputs.  The ${U}(\alpha,\beta)$ filters and maps resulting amplitudes onto the final qubits, which are then measured in the computational basis for SUM (XOR) and CARRY (AND) outputs.

\begin{equation}
{U}(\alpha,\beta)=
\left(
\begin{array}{cccc}
 A & B-F &   B+F & 0\\
 B+F & A &   B-F &0\\
 B-F & B+F & A & 0 \\
 0 & 0 & 0 & 1 \\
\end{array}
\right),
\label{eq:logicmatrix}
\end{equation}

where
\begin{equation}
    \begin{split}
    A=\frac{1}{3} \left(2 \cos \left(\frac{2}{3} \pi  (\alpha +\beta )\right)+1\right),
\\ 
B=\frac{1}{3} \left(1-\cos \left(\frac{2}{3} \pi  (\alpha +\beta )\right)\right),\\
F=\frac{\sin \left(\frac{2}{3} \pi  (\alpha +\beta )\right)}{\sqrt{3}}.
\end{split}
\label{eq:logicparameters}
\end{equation}

The corresponding half-adder Hamiltonian is
\begin{equation}
H_\mathrm{ha}(\alpha,\beta)=\frac{2 i \pi  (\alpha +\beta )}{3 \sqrt{3}} h_\mathrm{ha}=\frac{2 i \pi  (\alpha +\beta )}{3 \sqrt{3}}\left(
\begin{array}{cccc}
 0 & -1 & 1 & 0 \\
 1 & 0 & -1 & 0 \\
 -1 & 1 & 0 & 0 \\
 0 & 0 & 0 & 0 \\
\end{array}
\right).
\label{eq:hamhalf}
\end{equation}
The time-evolution of this Hamiltonian $U(\alpha,\beta) = \exp(-ih_{ha}t_\mathrm{eff}(\alpha, \beta))$ shows that the solution of the half-adder can be extracted on the timings given by $t_\mathrm{eff}(\alpha,\beta )=\frac{2 i \pi  (\alpha+\beta )}{3 \sqrt{3}}$. Throughout this work, we adopt units where the reduced Planck constant $\hbar = 1$.

Given the initial state $\ket{00}=\left(
\begin{array}{cccc}
 1 &
 0 &
 0 &
 0 
\end{array}
\right)^T$
we can fulfill the half-adder truth table and response as:

\begin{align}
\begin{split}
    U(0,0)\ket{00}=\ket{00}\\
    U(0,1)\ket{00}=\ket{01}\\
    U(1,0)\ket{00}=\ket{01}\\ 
    U(1,1)\ket{00}=\ket{10}
     \end{split}
 \label{eq:trut}
 \end{align}

We can generalize now the protocol with full-adder as depicted in Fig.~\ref{fig:2qubit_halfadder} (b). Again the device employs two qubits prepared in $\ket{\psi(t=0)}=\ket{00}$.  A 4\,\(\times\)\,4 unitary ${U}(\alpha, \gamma,\beta)$ confines dynamics to a protected sub‑manifold and %rejects leakage which e
encodes the angles $(\alpha, \gamma,\beta)\in\{0,1\}^2$ as logical inputs.  The ${U}(\alpha, \gamma,\beta)$ filters and maps the resulting amplitudes to the final qubits, which are then measured in the computational basis for the SUM (XOR) and CARRY outputs.

The corresponding matrix has the following form:

%\begin{widetext}

\begin{equation}
\scalebox{0.83}{$
 {U}(\alpha,\gamma,\beta)= \frac{1}{4}  \left(
\begin{array}{cccc}
 l+2 m+1 & -2 n-p-i q+1 & l-2 m+1 & 2 n-p-i q+1 \\
 2 n-p-i q+1 & l+2 m+1 & -2 n-p-i q+1 & l-2 m+1 \\
 l-2 m+1 & 2 n-p-i q+1 & l+2 m+1 & -2 n-p-i q+1 \\
 -2 n-p-i q+1 & l-2 m+1 & 2 n-p-i q+1 & l+2 m+1 \\
\end{array}
\right)$},
\label{eq:fulladdermatrix}
\end{equation}
%\end{widetext}

where

\begin{equation}
    \begin{split}
        l=e^{i \pi  (\alpha +\beta +\gamma )},\\
        m=\cos \left(\frac{1}{2} \pi  (\alpha +\beta +\gamma )\right),\\
        n=\sin \left(\frac{1}{2} \pi  (\alpha +\beta +\gamma )\right),\\
        p=\cos (\pi  (\alpha +\beta +\gamma )),\\
        q=\sin (\pi  (\alpha +\beta +\gamma )).
    \end{split}
\end{equation}
See Sec.~\ref{appendix} Appendix for the corresponding steps to arrive at Eq.~(\ref{eq:fulladdermatrix}).

The corresponding full-adder Hamiltonian reads

\begin{equation}
    H_{fa}=\frac{1}{4} \pi  (\alpha +\beta +\gamma ) h_{fa} = \frac{1}{4} \pi  (\alpha +\beta +\gamma ) \left(
\begin{array}{cccc}
 -1 & 1-i & -1 & 1+i \\
 1+i & -1 & 1-i & -1 \\
 -1 & 1+i & -1 & 1-i \\
 1-i & -1 & 1+i & -1 \\
\end{array}
\right).
\label{eq:hamfull}
\end{equation}
The time-evolution of this Hamiltonian $U(\alpha,\beta) = \exp(-ih_{fa}t_\mathrm{eff}(\alpha, \beta,\gamma))$ shows that the solution of the full-adder can be extracted on the timings given by $t_\mathrm{eff}(\alpha , \beta,\gamma)=\frac{1}{4} \pi  (\alpha +\beta +\gamma )$.

Applying the full-adder gate we receive the corresponding truth table

\begin{align}
\begin{split}
    {U}(0,0,0)\ket{00}=\ket{00}&\quad
    {U}(0,0,1)\ket{00}=\ket{01}\\
  {U}(0,1,0)\ket{00}=\ket{01}&\quad
    {U}(0,1,1)\ket{00}=\ket{10}\\
    {U}(1,0,0)\ket{00}=\ket{01}&\quad
    {U}(1,0,1)\ket{00}=\ket{10}\\
    {U}(1,1,0)\ket{00}=\ket{10}&\quad
    {U}(1,1,1)\ket{00}=\ket{11}.
\end{split}
    \label{eq:truthtable2}
\end{align}

\section{Conclusion}
In this study, we presented Quantum Hamiltonian Computing as a method for performing quantum arithmetic operations, which reduces qubit overhead by encoding classical inputs within a single quantum gate. Our two-qubit half- and full-adder circuits in a 4×4 Hilbert space require fewer resources than traditional textbook reversible design built from a Toffoli plus clean-up gates \cite{cuccaro2004new,Vedral1996,amy2013, Selinger2013}. Leveraging unitary evolution enables us to bypass classical CMOS energy limitations and achieve reversible logic with fewer qubits and gates, where the qubit count is only logarithmic in dependence on possible output states. Furthermore, the QHC unitary realizes the complete truth-table in one analogue pulse, as long as the gate can be implemented on the corresponding quantum hardware.
While the current design targets classical logic without utilizing superposition of input and output states, it should be clear that this is possible in principle and can be as simple as converting the inputs from binary to real value inputs. However, the found unitaries are not fully expressive but can only map a sub-manifold in the Hilbert space. Removing this limitation, would lead again to scratch qubits thus no benefit to previous implementations \cite{Vedral1996}. However, the sub-manifold can be optimized to evaluate to superposition states expected to become most prominent in the circuit without the need to introduce qubits.

The QHC framework shows promise for advancement of FPGA designs by taking advantage of quantum architectures to minimize energy consumption. This approach offers potential applications in quantum circuits, integrated quantum photonics, and other emerging technologies, especially as the operational speeds of integrated quantum photonic devices, for example, can exceed those of classical computers \cite{PRXQuantum.5.010101}. 

 \section*{Acknowledgments}
 
We would like to thank the Research Initiative ‘Quantum Computing for Artificial Intelligence’ (QC-AI) for their support.

\section{Appendix: Full-adder unitary matrix ${U}(\alpha,\gamma,\beta)$  construction}
\label{appendix}
Let the computational basis of two qubits be ordered as
\(\{\ket{00},\ket{01},\ket{10},\ket{11}\}\).
We could fix the column vector (the \emph{control} or \emph{eigenvector}) as:
\begin{equation}
  V \;=\; \ket{00},
  \tag{A.1}
\end{equation}

and three real parameters \(\alpha,\gamma,\beta\in\mathbb R\).
The task is to build a \(4\times4\) unitary matrix
\({U}(\alpha,\gamma,\beta)\) such that, when each parameter is restricted
to the Boolean set \(\{0,1\}\), the eight input triples are mapped as

\begin{center}
\begin{tabular}{c|c}
$(\alpha,\!\gamma,\!\beta)$ & $U(\alpha,\gamma,\beta)\,V$ \\ \hline
$(0,0,0)$ & $\ket{00}$ \\
$(0,0,1)$ & $\ket{01}$ \\
$(0,1,0)$ & $\ket{01}$ \\
$(0,1,1)$ & $\ket{10}$ \\
$(1,0,0)$ & $\ket{01}$ \\
$(1,0,1)$ & $\ket{10}$ \\
$(1,1,0)$ & $\ket{10}$ \\
$(1,1,1)$ & $\ket{11}$
\end{tabular}
\end{center}

We define now a $4$-cycle permutation matrix $\mathcal R$ as

\begin{equation}
  \mathcal R \;=\;
  \ket{01}\bra{00}\;+\;
  \ket{10}\bra{01}\;+\;
  \ket{11}\bra{10}\;+\;
  \ket{00}\bra{11}.
  \tag{A.2}
\end{equation}
Equation (A.2) permutes the basis cyclically
\(\ket{00}\!\to\!\ket{01}\!\to\!\ket{10}
\!\to\!\ket{11}\!\to\!\ket{00}\); hence
\(\mathcal R^{4}=I_{4}\).
Written as a matrix,
\[
  \mathcal R
  =\begin{pmatrix}
     0&0&0&1\\
     1&0&0&0\\
     0&1&0&0\\
     0&0&1&0
   \end{pmatrix},
   \qquad
   \mathcal R^\dagger\mathcal R=I_{4}.
\]

Because \(\mathcal R\) is unitary, the principal matrix logarithm
\begin{equation}
  \mathcal R = e^{-iH},
  \qquad
  H \;=\; i\,\mathrm{log}\,\mathcal R,
  \qquad
  H^\dagger = H,
  \tag{A.3}
\end{equation}
is Hermitian.  (The eigenvalues of \(\mathcal R\) are the fourth roots of
unity \(\{1,i,-1,-i\}\).)

the three–parameter family could be Defined as
\begin{equation}
  %\boxed{%
    U(\alpha,\gamma,\beta)
      \;=\;
      \exp\!\bigl[-\,i\,(\alpha+\gamma+\beta)\,H\bigr]
      \;\;
      %}.
  \tag{A.4}
\end{equation}

Some algebra leads to Eq.~(\ref{eq:fulladdermatrix}).

\end{document}